\documentstyle[preprint,revtex]{aps}
\begin{document}
\hfill\vbox{\hbox{\bf NUHEP-TH-93-19}\hbox{July 1993}}\par
\thispagestyle{empty}
\begin{title}
\begin{center}
{\bf $B_c$ Mesons Production at Hadron Colliders by \\
Heavy Quark Fragmentation}
\end{center}
\end{title}
\author{Kingman~Cheung}
\begin{instit}
Dept. of Physics \& Astronomy, Northwestern University, Evanston,
Illinois 60208, USA\\
\end{instit}
\begin{abstract}
\nonum
\section{Abstract}
We present a reliable estimate on the production rate of $B_c$ mesons
in $1S$ and $2S$ states in
the large transverse momentum region at hadronic colliders
using heavy quark fragmentation
functions derived within the framework of perturbative QCD.  We
also present the transverse momentum  distribution for the $B_c$ mesons.
The production rate is large enough for the $B_c$ mesons to be identified at
the Tevatron.  At the SSC/LHC the rate is so large that their properties can be
studied in details.
\end{abstract}

\section{Introduction}
\label{intro}

Heavy flavor production and decays are very useful for
 measuring CKM matrix elements of the standard model and for
testing bound-state models for mesons and
baryons.  Ever since the first $B$ meson was discovered, a lot of data have
been coming out from, {\it e.g.},
CLEO, ARGUS, LEP/SLC, FNAL, on the $B_d$ and $B_u$ meson families.  From
the recent run at Tevatron the masses and other properties
of the $B_s$ have been measured and confirmed \cite{Bs}.
The next family of  $B$ mesons will be the $B_c$ mesons
 made up of $\bar bc$.
The $B_c$-meson family
differs from the $J/\psi$ and  $\Upsilon$ families and from other $B$ mesons
because it is made up of a pair of heavy quark and antiquark  of {\it
different} flavors and masses.
The $J/\psi$ and $\Upsilon$ families have
played  important roles in developing heavy quark
bound-state models inspired by QCD.
Being  quarkonium systems of different flavors and  masses
$\bar b c$ bound-states
provide unique opportunities to test different bound-state
models of QCD.  The decays of $B_c$ mesons also provide rich sources to
test the standard model, e.g. the measurement of $|V_{bc}|$, and enables
us to see the interplay between  strong and weak interactions.

Since the physics of the $B_c$ mesons is so interesting, one would like to know
how many can be produced in the present colliders (e.g. Tevatron) and in
the future hadronic supercolliders (SSC/LHC).  It is the purpose of this
letter to present reliable estimate in the  high $p_T$ region,
by using the heavy
quark fragmentation functions $D_{\bar b \to B_c}(z)$ \cite{ours}
which are  based on
perturbative QCD.  We will summarize some features of these heavy quark
fragmentations in Sec.~II, and present the production rates and the $p_T$
distributions for the $B_c$ mesons in Sec.~III.
Finally we will conclude in Sec.IV.

\section{Heavy Quark Fragmentation}

Previous estimates of $B_c$ meson production have based
on perturbative QCD calculations for $e^+e^-$ colliders \cite{bclep}
and monte carlo studies for both $e^+e^-$ and hadronic colliders \cite{bcmc}.
The monte carlo  estimates of the ratio $\sigma(B_c^\pm)/\sigma(b\bar b X)$
are all of the
order $10^{-3}$ for LEP, Tevatron, SSC/LHC, and HERA.  This fact leads
us to think that the production mechanisms are all of the same nature.
In the region of large $p_T$, the major mechanism for  producing $B_c$ or any
other mesons is heavy quark fragmentation \cite{ours,eric},
in which a $\bar b$  antiquark
is produced at large $p_T$ by a hard-scattering process and it subsequently
fragments into the  meson.
The differential cross section for direct production of the
$B_c$ meson at high $p_T$ can be factorized at leading order in $\alpha_s$ as
\begin{equation}
\label{fact}
d\sigma(B_c(p)) = \int_0^1 dz\; d\hat\sigma(\bar b(p/z,\,\mu))
D_{\bar b\to B_c}(z,\,\mu)\,,
\end{equation}
where $z$ is the longitudinal momentum fraction carried by the $B_c$,
and $\mu$ is a factorization scale.
The physical interpretation is as follow: a heavy $\bar b$ antiquark is
produced in a hard process with four-momentum $p/z$, and then it fragments
into the  $B_c$ meson with a longitudinal momentum fraction $z$.
The fragmentation function $D_{\bar b\to B_c}(z)$ satisfies the
Altarelli-Parisi evolution equation
\begin{equation}
\label{evol}
\mu \frac{\partial}{\partial \mu} D_{\bar b\to B_c}(z) = \int_z^1
\frac{dy}{y} P_{\bar b \to \bar b}(z/y, \mu) D_{\bar b\to B_c}(y,\mu)\,,
\end{equation}
at leading order in $\alpha_s$.
The factorization for $B_u,\,B_d$, and $B_s$ productions can be described
in the same way as Eqs.~(\ref{fact}) and (\ref{evol}), with the corresponding
fragmentation functions.
These fragmentation functions should be independent of the
hard process by which the $\bar b$ is  produced.

The fragmentation of $\bar b$ into $B_u,\,B_d$, and $B_s$ is a soft process,
and can only be described by a phenomenological function \cite{peter}.
However, $\bar b\to B_c$ requires the production of a $c\bar c$ pair and it is
therefore a hard process which can be calculated using perturbative QCD
\cite{ours,eric}.  This  perturbative QCD approach has been shown valid in
calculating the fragmentation functions for heavy quarkonium productions
 \cite{eric},
including the splitting of gluons and charm quarks into $S$-wave charmonium.
The fragmentation functions $D_{\bar b\to B_c}(z)$ derived in
Ref.~\cite{ours} need only the input parameters of
 $\alpha_s,\,m_b,\,m_c$, and the radial wave function $R(0)$ of the bound
state at the origin so that it has more predictive power.
The  initial fragmentation functions are given by
\begin{eqnarray}
D_{\bar b\rightarrow B_c}(z,\mu_0) & = &
\frac{2\alpha_s(2m_c)^2 |R(0)|^2}
{81\pi m_c^3}\; \frac{rz(1-z)^6}{(1-(1-r)z)^6} \nonumber \\
\label{dz1}
&\times & [ 6 - 18(1-2r)z + (21 -74r+68r^2) z^2 \\
 && -2(1-r)(6-19r+18r^2)z^3  + 3(1-r)^2(1-2r+2r^2)z^4 ]\,,  \nonumber
\end{eqnarray}
for the $^1S_0$ $B_c$ state, and
\begin{eqnarray}
D_{\bar b\rightarrow B_c^*}(z,\mu_0) & = &
\frac{2\alpha_s(2m_c)^2 |R(0)|^2}
{27\pi m_c^3}\; \frac{rz(1-z)^6}{(1-(1-r)z)^6} \nonumber \\
\label{dz2}
&\times & [ 2 - 2(3-2r)z + 3(3 - 2r+ 4r^2) z^2 \\
& &   -2(1-r)(4-r +2r^2)z^3  + (1-r)^2(3-2r+2r^2)z^4 ]\,,  \nonumber
\end{eqnarray}
for the first excited  state $B_c^*$ which has the spin-orbital quantum number
$^3S_1$, and $r=m_c/(m_b+m_c)$.  Note that the scale
in $\alpha_s$ is set to be $2m_c$.
The perturbative QCD calculation gives directly the fragmentation function at
the scale $\mu_0=2m_c$, which is the minimum virtuality of  the
gluon splitting into $c\bar c$.
  Using the Heavy Quark Effective Theory methods \cite{jaffe},
it can be shown that the evolution up to the scale $m_b$ is trivial, which
means that higher order radiative corrections will not introduce any logarithm
of $m_b/m_c$.  A
convenient choice for the initial scale is $\mu_0=m_b+2m_c$ \cite{ours},
because the
fragmentation functions for $c\to B_c,\,B_c^*$ can then be obtained from
Eqs.~(\ref{dz1}) and (\ref{dz2}) simply by interchanging $m_b$ and $m_c$.

In the factorization scheme of Eq.~(\ref{fact}), all the dependence on the
momentum $p$ is in the hard process $\hat\sigma$.
Large logarithm of $p/\mu$ can be avoided by choosing the factorization scale
$\mu$ to be of order $p$.
The induced large logarithm of order $\mu/m_b$ in $D(z)$ can be solved by
evolving the Eq.~(\ref{evol}).
To leading order in $\alpha_s$ only the $\bar b \to \bar b$ contributes to
the evolution. The fragmentation functions $D_{\bar b\to B_c}(z)$ and
 $D_{\bar b\to B_c^*}(z)$ at the initial scale $\mu_0$ and higher scale
have been shown in Fig.~1 of Ref.~\cite{ours}.

\section{Results \& Discussions}

In this section, we use Eq.~(\ref{fact}) to compute the direct production
rates of $B_c$ and $B_c^*$ in hadronic collisions.
Our calculation  includes
 \begin{equation}
 gg  \to  b\bar b\,,\quad  g \bar b  \to g\bar b\,,\quad {\rm and}\quad
 q\bar q  \to  b \bar b \,,
\end{equation}
as the hard subprocesses for the inclusive production of the $\bar b$.  We
choose the scale $\mu$ for the parton distribution functions and for
$\alpha_s$ to be  the transverse mass of the $\bar b$,
$\sqrt{p_{T\bar b}^2 + m_b^2}$.
We use the parameterization of  HMRS (set b) \cite{HMRS}
for parton distribution functions.
The running coupling constant $\alpha_s(Q)$ is evaluated at 1-loop by
evolving from the
experimental value $\alpha_s(m_Z)=0.12$ \cite{zlep}, and given by
\begin{equation}
\alpha_s(Q) = \frac{\alpha_s(m_Z)}{1+ 8\pi b_0\alpha_s(m_Z)\log(Q/m_Z)}\,,
\end{equation}
where $b_0 = (33-2n_f)/48\pi^2$, and $n_f$ is the number of active flavors
below the scale $Q$.
The subprocess cross sections are convoluted with $D(z,\mu)$, as is in
Eq.~(\ref{fact}).  The functions $D(z,\mu_0)$ at the initial scale $\mu_0$ are
given in Eqs.~(\ref{dz1}) and (\ref{dz2}), and are evolved
to the scale
$\mu$ using Eq.~(\ref{evol}).
For the initial fragmentation functions $D(z,\mu_0)$
we use the input parameters of $m_b=4.9$~GeV, $m_c=1.5$~GeV, and
$|R(0)|^2=(1.18\;{\rm GeV})^3$.

The $p_T$ spectrum for the $B_c$ meson at
Tevatron energy is shown  in Fig.~\ref{teva}, with the acceptance cuts
\begin{equation}
 p_T(B_c) > 10 \;{\rm GeV}\quad {\rm and }\quad |y(B_c)| < 1\,.
\end{equation}
The corresponding spectrum for the $B_c^*$ is also shown on the same figure.
The shape of the two spectra is very similar, because
$D_{\bar b\to B_c}(z)$ and $D_{\bar b\to B_c^*}(z)$ have similar shapes and
differ primarily by an overall normalization difference of about 50\%.
The corresponding spectra at SSC($\sqrt{s}=40$~TeV) and LHC($\sqrt{s}=14$~TeV)
 energies are shown in  Fig.~\ref{SSC}, but under  slightly different
acceptance requirements
\begin{equation}
 p_T(B_c) > 20 \;{\rm GeV}\quad {\rm and }\quad |y(B_c)| < 2.5\,.
\end{equation}
The integrated cross sections versus $p_T^{\rm min}(B_c)$ are  also
shown in  Fig.~\ref{inte}.
The cross sections  at the SSC are about three times as large as those at the
LHC, and about two order of magnitudes larger than those at the Tevatron.

So far we have only estimated the $B_c$ meson productions in $1\,^1S_0$ and
$1\,^3S_1$ states.
Since the annihilation channel for the decay of the excited $B_c$-meson
 states is  suppressed relative to the
electromagnetic and hadronic transitions, all the
excited states ($1\,^3S_1,\,2S,\,1P,\,2P,\,1D$) below the $BD$ threshold will
decay to the ground state $1\,^1S_0$ by emitting  photons or pions.  They
therefore all contribute to the inclusive production of $B_c$ mesons.
A simple modification can be made to estimate the
productions in $2\,^1S_0$ and $2\,^3S_1$ states, by multiplying with the factor
$|R_{2S}(0)/R_{1S}(0)|^2 \simeq 0.6$ \cite{eichten}.  Therefore, the curves in
Figs.~\ref{teva}, \ref{SSC}, and \ref{inte} can be
multiplied by 0.6 to get the productions for 2S states.
To get
the total inclusive production rate, however, we need to include $P$-wave and
possibly $D$-wave contributions, which have not been calculated.  Therefore,
the cross sections  presented here are rather conservative to the actual
production rates.
Table~\ref{table1} shows the number of $B_c$
mesons  that can be produced inclusively  at Tevatron, SSC and LHC, including
the contributions from $1S$ and $2S$ states, with
integrated luminosities of 0.025, 10, and 100
fb$^{-1}$ respectively.
It is also informative to give the ratio $\sigma(B_c)/\sigma(\bar b X)$,
which is simply the fragmentation probability $\int_0^1 dz D_{\bar b\to
B_c}(z)$.  Adding the contributions from $1S$ and $2S$ states, the ratio
$\sigma(B_c)/\sigma(\bar b X)$ is about $1.5\times 10^{-3}$, which is
consistent with the monte carlo studies \cite{bcmc}.

The factorization in Eq.~({\ref{fact}) is correct up to the
order $(m_b/p_T)^2$, which explains why we impose a rather high $p_T$ cut on
the $B_c$ mesons.
In the low $p_T$ region, mechanisms other than heavy quark
fragmentation have to be taken into account.  For example, production of pairs
of $b\bar b$ and $c\bar c$ followed by recombination of $\bar b$ and $c$ to
form a $B_c$ meson contributes  at low $p_T$ region; whereas heavy quark
fragmentation dominates at high $p_T$.
Similar conclusion can be found in the production of $J/\psi$ by
heavy quark fragmentation \cite{sean}, which is dominant over
the process $gg\to\psi g$ in the large $p_T$ region.
There are other uncertainties arising from higher order QCD corrections,
relativistic corrections of the bound state model, and the values of
$\alpha_s$ and $R(0)$ used.  But the largest source of uncertainties comes from
the values of $m_b$ and $m_c$ employed
because of the factor $1/m_c^3$ in the
initial fragmentation functions as in Eqs.~(\ref{dz1}) and (\ref{dz2}).

\section{Conclusions}

We have used heavy quark fragmentation functions derived from perturbative QCD
to calculate the production rates and the $p_T$ distributions for
$S$-wave $B_c$ mesons at Tevatron, SSC , and LHC energies.
Imposing cuts of  $p_T(B_c)>10$~GeV and $|y(B_c)|<1$  at the Tevatron,
and including the
contributions from 1S and 2S states, about 16000 $B_c^+$ and 16000 $B_c^-$
mesons
should be produced for 25 pb$^{-1}$ integrated luminosities.
The corresponding numbers for SSC and LHC with $p_T(B_c)>20$~GeV and
$|y(B_c)|<2.5$
are $7.1\times 10^7$  and $2.3\times 10^8$
 with  integrated luminosities of 10 and 100 fb$^{-1}$ respectively.
These  $B_c$ mesons can be detected via the decays of the form
$J/\psi +  X$, and in particular, via
\begin{equation}
B_c^\pm \quad \rightarrow  \psi  \ell^\pm \nu_\ell \quad {\rm and} \quad
B_c^\pm \quad \rightarrow  \psi   \pi^\pm
\end{equation}
with $\psi \rightarrow  \ell'^+ \ell'^-$.
The first one has a distinct signature of three charged leptons coming off from
the same secondary vertex and  has a combined branching ratio of
about 1\%.   The second decay channel has a smaller  branching ratio of order
0.2-0.4\%, but it has the advantage that the $B_c$ can be fully reconstructed.
The production rate of $B_c$ mesons given above is large enough that it should
be possible to use these decay modes to identify the $B_c$ mesons at the
Tevatron, and to study its properties in details at the SSC and LHC.

\acknowledgements

Special thanks to Eric Braaten for suggesting the problem, reading the
manuscript, and giving many
valuable comments.  Also thanks to E.~Eichten, R.~Oakes and T.~C.~Yuan for
helpful discussions.
This work was supported by the U.~S. Department of Energy, Division of
High Energy Physics, under Grant DE-FG02-91-ER40684.


\begin{table}
\caption{\label{table1}
Table showing the number of $B_c$ mesons that can be produced at Tevatron,
SSC, and LHC,  including the contributions from $1S$ and $2S$ states, and
under different $p_T^{\rm min}(B_c)$ cuts.
Another acceptance cut is $|y(B_c)|<1\;(2.5)$ at Tevatron (SSC/LHC).
The   integrated
luminosities are  25 pb$^{-1}$, 10 fb$^{-1}$ and 100 fb$^{-1}$ for the
Tevatron, SSC, and LHC respectively.
}
\begin{tabular}{c|ccc}
$p_T^{\rm min}(B_c)$  & Tevatron & SSC  & LHC \\
(GeV) & & & \\
\hline
10                   &   16000  &  --  & --  \\
15                   &   4400   &  --  & --  \\
20      & 1600       &  7.1$\times 10^7$  &  2.3$\times 10^8$  \\
30      & 300        &  2.3$\times 10^7$  &  6.6$\times 10^7$  \\
40      & 82         &  9.8$\times 10^6$  &  2.5$\times 10^7$   \\
50      & 28         &  4.9$\times 10^6$  &  1.2$\times 10^7$
\end{tabular}
\end{table}
\newpage
\figure{\label{teva}
The dependence of the differential cross section $d\sigma/dp_T(B_c)$
on the transverse momentum $p_T(B_c)$ for  the
ground state $B_c(1\,^1S_0)$ and the first excited
state $B_c^*(1\,^3S_1)$ at the Tevatron.}

\figure{\label{SSC}
The dependence of the differential cross section $d\sigma/dp_T(B_c)$
on the transverse momentum $p_T(B_c)$ for
the ground state $B_c(1\,^1S_0)$ and the first excited
state $B_c^*(1\,^3S_1)$ at the SSC and LHC.}

\figure{\label{inte}
The integrated cross sections $\sigma \left ( p_T(Bc) > p_T^{\rm min}(Bc)
\right)$  for the ground state $B_c(1\,^1S_0)$ (solid) and the first
excited state $B_c^*(1\,^3S_1)$
(dash) versus  $p_T^{\rm min}(B_c)$  at the Tevatron, SSC, and LHC.}

\end{document}